\documentclass[showpacs,10pt]{iopart}
\usepackage{color}
\usepackage{times}
\usepackage{iopams}
\usepackage{cite}
\usepackage{graphicx} 

\newcommand{\rC}{\textbf{C}}
\newcommand{\rI}{\textbf{I}}
\newcommand{\x}{{\mathbf x}}

\newcommand{\ecut}{\epsilon_{\rm{cut}}}

\newcommand{\EQ}[1]{\begin{eqnarray}#1\end{eqnarray}}

\begin{document}
\title{Stochastic longevity of a dark soliton in a finite-temperature Bose-Einstein condensate}
\author{K. J. Wright and A. S. Bradley} 
\address{Jack Dodd Center for Quantum Technology, Department of Physics, University of Otago, Dunedin, New Zealand.}
\begin{abstract}
We study the decay of a dark soliton in a homogeneous Bose-Einstein condensate. We give an analytical treatment of a decaying soliton, deriving an expression for the soliton velocity and decay time in the absence of thermal noise. We test the result against numerical simulations of a spatially confined system and find good agreement in the regime of low temperature ($k_BT\ll \mu$). Thermal fluctuations are found to slow the escape of the soliton, extending its lifetime beyond the predictions of the noise-free theory; the effect becomes significant at a characteristic temperature $k_BT\sim \mu$. This stabilization by noise allows us to infer an analytical lower bound for the dark soliton decay time. 
\end{abstract}
\section{Introduction}
The observation of dark~\cite{Burger99a} and bright~\cite{Strecker2002a} solitons in Bose-Einstein condensates (BEC) has sparked much renewed interest in nonlinear excitations of one-dimensional (1D) superfluids. As localized translating solutions of the repulsive, completely integrable nonlinear Schr\"{o}dinger equation~\cite{Kivshar1998a,Frantzeskakis10a}, dark solitons also form long-lived excitations in trapped, finite-temperature Bose-Einstein condensates. Unlike quantum vortices~\cite{Fetter2001}, dark solitons in simply connected geometries lack topological protection and thus the effect of thermal fluctuations on their stability is of particular interest. They have been created through a variety of experiments, including phase imprinting~\cite{Burger99a,Denschlag2000a,Anderson2001b,Becker08a,Stellmer08a}, using slow light~\cite{Dutton2001a}, shock waves~ \cite{Engels07a}, by merging independent BECs~\cite{Weller08a}, or through superfluid counterflow~\cite{Hamner11a}. Meanwhile, theoretical effort has focused on disturbing the ideal homogeneous soliton with thermal~\cite{Fedichev1999b,Busch2000,Jackson07a,Cockburn10a,Martin2010b} and quantum~\cite{Dziarmaga2003a,Dziarmaga04a,Dziarmaga2006a,Mishmash2009a,Mishmash2009b,Dziarmaga2010a,Sykes2011a} fluctuations, or by breaking integrability~\cite{Muryshev2002a} with a disordered potential~\cite{Bilas2005a}, a stochastic perturbation~\cite{Carr2001a}, by parametrically driving the soliton~\cite{Proukakis2004a}, or confining the 1D system in a non-homogeneous external potential~\cite{Parker2003,Konotop2004a,Brazhnyi2006a}. 

Significant theoretical progress has been made in the low temperature regime. In the zero temperature limit a Bogolyubov analysis of the breakdown of integrability due to weak transverse confinement predicted lifetimes very similar to experimental observations~\cite{Muryshev2002a}. Numerical simulations based on a two-fluid dynamical model of the BEC coupled to a thermal component~\cite{Jackson07a} were able to reproduce the general decay phenomena observed at relatively low temperature~\cite{Burger99a}. More recently, application of Hamiltonian truncated Wigner theory revealed that fluctuations can slow the soliton below its classically predicted speed~\cite{Martin2010a,Martin2010b}. In a grand-canonical treatment of the trapped Bose gas~\cite{Cockburn10a}, soliton decay according to the stochastic Gross-Pitaevskii equation (SGPE)~\cite{Stoof1999} was well approximated by applying perturbation theory to the purely damped GPE~\cite{Pitaevskii1958}, provided the system temperature is much lower than the energy scale of the soliton. This progress highlights the need for studies at higher temperatures, where thermal fluctuations can significantly modify the mean-field dynamics of the dark soliton. 

The pervasiveness of thermal fluctuations and their importance in 1D ultra-cold gas systems motivates our study of their role in the homogeneous dark soliton system. In particular, we are interested in the effect of thermal fluctuations when the system temperature is comparable to the soliton energy, and the departure of soliton motion from the predictions of perturbation theory. In the harmonically trapped system~\cite{Martin2010a,Cockburn10a} the soliton moves through a wide range of background particle densities. This complicates the analysis as the net effect of thermal fluctuations depends on the local background density of the condensate. By restricting our attention to the homogeneous system we aim to obtain a clear point of comparison for the effects of fluctuations. In addition to being simpler to analyze, the homogeneous regime is experimentally realizable in modern highly flexible experimental setups~\cite{Arnold06a,Schnelle08a,Henderson09a,Sherlock11a}. 

We apply Lagrangian perturbation theory~\cite{Kivshar1995a,Kivshar1994a,Kivshar1989,Theocharis2005,Kong2010,Cockburn10a} to find an expression for the soliton velocity throughout its decay in an infinite homogeneous system. This provides a simple quantitative basis for the intuitive picture of dark soliton decay in BECs, and a baseline for our grand-canonical c-field treatment based on the stochastic projected Gross-Pitaevskii equation (SPGPE)~\cite{SGPEI,SGPEII,Blakie08a}. Comparing our analytical result with SPGPE simulations of a large finite system, we find excellent agreement at low temperature where damping dominates over noise. Thermal fluctuations in the SPGPE become important with increasing temperature, and their main effect is to slow the soliton relative to damped GPE theory, in clear departure from our analytical predictions. We thus find an expression for the decay time of a dark soliton that serves as a useful lower bound for the lifetime in the grand-canonical c-field theory.
 
The structure of this article is as follows. In the remainder of this section we summarize the SPGPE theory for quasi-equilibrium systems, and the reduction to a 1D effective theory. In Sec.~\ref{LPT} we apply Lagrangian perturbation theory to the dark soliton in a homogeneous background, finding analytical expressions characterizing the decaying dark soliton. In Sec.~\ref{SIMS} we present numerical simulations of the dark soliton evolution, comparing our analytical result with the results of SPGPE and damped GPE numerical simulations.

\subsection{Stochastic projected Gross-Pitaevskii formalism}
The SPGPE provides a grand-canonical theory of finite-temperature BECs based on the truncated Wigner approxmation~\cite{SGPEI,SGPEII,Bradley2005b,Bradley08a,Blakie08a}~(see also the path-integral treatment~\cite{Stoof1999,Stoof2001,Proukakis06a,Proukakis08a,Cockburn09a}). In addition to the usual GPE terms associated with Hamiltonian evolution, the equation of motion contains damping and noise terms describing interactions of low-energy atoms with a high-energy thermal reservoir. By incorporating the energy cutoff defining the thermal reservoir explicitly, the stochastic PGPE has the advantage that it may be consistently formulated in 1, 2 and 3 spatial dimensions, is able to quantitatively model experiments, and is valid right across the phase transition~\cite{Weiler08a}.

A finite temperature, partially condensed Bose gas consists of a BEC, a number of low-energy excitations, and a large number of high-energy modes that are weakly occupied. To arrive at the SPGPE, the low-energy region beneath an energy cutoff $\ecut$ is described by a classical field  ($\rC$-field) theory~\cite{Blakie08a}, and we refer to this region as the \emph{coherent} region.
 The \emph{incoherent} ($\rI$) region of phase-space (above $\ecut$) is coupled to the coherent region via non-Hamiltonian two-particle scattering processes which take the form of damping and noise terms in the equation of motion.  
Neglecting the less important particle conserving processes (scattering terms of Ref.~\cite{SGPEII} which are small near equilibrium), the SPGPE reduces to the simple growth SPGPE~\cite{Bradley08a,Blakie08a}
\EQ{\label{GrowthSGPE}
d\psi(\x,t)&=&{\cal P}\left\{\frac{1}{\hbar}(i-\gamma)\left(\mu-L_{GP}\right)\psi(\x,t) dt+dW_\gamma(\x,t)\right\}.
}
Here the GP-operator is $L_{GP}\psi\equiv(-\hbar^2\nabla^2/2m+V_{\rm ext}(\x)+U_0|\psi|^2)\psi$, and $U_0=4\pi\hbar^2 a/m$ for s-wave scattering length $a$. The projection operator ${\cal P}$ imposes an energy cutoff, formally eliminating modes above $\ecut$ as part of a thermal reservoir described by $\mu$, and $T$. The noise is a complex-gaussian Wiener-process,with non-vanishing correlation function,
\EQ{\label{fulldW}
\langle dW^*_\gamma (\x,t)dW_\gamma (\x',t')\rangle&=&\frac{2\gamma k_B T}{\hbar}\delta(\x-\x')\delta(t-t')dt.
}
This form of the noise is obtained from a first-principles microscopic treatment of reservoir interactions~\cite{SGPEII,Bradley08a}, and ensures that equilibrium ensemble properties are independent of $\gamma$.

The dimensionless dissipation rate of the stochastic-PGPE theory can be calculated near equilibrium in 3D systems~\cite{Bradley08a}, and takes the form
\begin{eqnarray}\label{gamdef} \gamma&=&\gamma_0\sum_{k=1}^\infty\; \frac{e^{\beta\mu(k+1)}}{e^{2\beta\ecut k}}\Phi\left[\frac{e^{\beta\mu}}{e^{\beta\ecut}},1,k\right]^2,
\end{eqnarray}
where $\gamma_0=4ma^2k_BT/\pi\hbar^2$, and $\Phi$ is the Lerch transcendent. The remarkable feature of this expression is that it is independent of position. It is valid in the region $V_{\rm ext}(\x)\leq 2\ecut/3$~\cite{Bradley08a}, which typically encloses the majority of highly occupied modes, and becomes weakly spatially varying outside this region. The simplicity of this expression is a consequence of retaining the ultra-violet energy cutoff $\ecut$ in the theory\footnote{A consistent choice of the ultra-violet cutoff is typically of order $\ecut\lesssim 3\mu$~\cite{Blakie08a}.}. In general $\gamma$ is a small parameter depending on the system geometry, temperature, interaction strength, and particle number. The general phenomena considered in the present work require, as is always the case for the degenerate Bose gas, that $\gamma\ll 1$. 
\subsection{Effective 1D regime}\label{1dregime}
We consider a trap of the form $V_{\rm ext}(\x)=\frac{m}{2}\omega_\perp^2r_\perp^2+V(x)$ where $r_\perp=\sqrt{y^2+z^2}$.
We identify a 1D SPGPE regime defined by 
\EQ{
\frac{\hbar^2}{ma^2}&\ll\mu\lesssim\hbar\omega_\perp\lesssim k_BT.
}
The first condition ensures that scattering remains 3D, and the remaining conditions can be written in terms of length scales as
\EQ{
\sqrt{\frac{\mu}{m\omega_\perp^2}}\lesssim\sqrt{\frac{\hbar}{m\omega_\perp}}\lesssim\sqrt{ \frac{k_BT}{m\omega_\perp^2}},
}
which imposes a reduction of the BEC to 1D, while preserving the 3D character of the thermal cloud. We can thus obtain a 1D reduction of the SPGPE with effective interaction strength $g=2\hbar\omega_\perp a$, yet retain the damping rate given by the universal form \eref{gamdef}. 

Formally eliminating the transverse degrees of freedom, we obtain the 1D SPGPE
\EQ{\label{simpSGPE}
d\psi(x,t)&=&{\cal P}\left\{\frac{1}{\hbar}(i-\gamma)\left(\mu-L_{GP}\right)\psi(x,t) dt+dW_\gamma(x,t)\right\}.
}
where now $L_{GP}\psi\equiv(-\hbar^2\partial_x^2/2m+V(x)+g|\psi|^2)\psi$, and 
\EQ{
\langle dW^*_\gamma (x,t)dW_\gamma (x',t)\rangle&=&\frac{2\gamma k_B T}{\hbar}\delta(x-x')dt.
}

For our analytical treatment it is convenient to work in dimensionless units; we also set $V(x)\equiv 0$.
We neglect the projector (by setting ${\cal P}\equiv1$), and rescale in units of healing length $\xi^2=\hbar^2/m\mu$ and speed of sound $c=\sqrt{\mu/m}$, so that $\chi=x/\xi$, $\tau=ct/\xi$ are dimensionless space and time variables. Rescaling the wavefunction $\psi=\sqrt{n}u$, with background density $|\psi|^2=n=\mu/g$, we arrive at the dimensionless form of the SGPE
\EQ{\label{Perturbationequation}
i\frac{\partial u}{\partial \tau}+\frac{1}{2}\frac{\partial^2 u}{\partial \chi^2}-(|u|^2-1)u&=\gamma R(u),
}
with perturbation
\begin{equation}\label{Perturbationterm}
\gamma R(u)\equiv-i\gamma\left(-\frac{1}{2}\frac{\partial^2 u}{\partial \chi^2}+(|u|^2-1)u\right)+\xi(\chi,\tau).
\end{equation} 
The non-vanishing correlation of the Langevin noise is 
\EQ{\label{dimlessdw}
\langle\xi^*(\chi,\tau) \xi(\chi',\tau)\rangle&=&2\gamma\sigma \delta(\chi-\chi')\delta(\tau-\tau').
}
The size of the rescaled temperature $\sigma\equiv k_B T/\mu$ compared to unity thus plays a central role in any nonlinear phenomena for finite temperature atomic superfluids.

 \section{Dissipative dynamics}\label{LPT}
To gain some understand of the different aspects of dissipation in soliton decay, we first consider the evolution in the absence of thermal noise. Our primary concern then is to understand effect of the terms explicitly proportional to $\gamma$ in Eq.~(\ref{simpSGPE}) on a dark soliton that would otherwise propagate through a homogeneous background without change to its depth and velocity. 

The dynamical problem of the soliton motion may be solved by using a collective variable ansatz for the wavefunction. Assuming a specific form for the wavefunction $u(\chi,a_1,\dots,a_n)$, where the collective variables $a_i$ are functions of time only, we minimize the action within the subspace of such wavefunctions. This produces a set of Euler-Lagrange equations for the variables $a_i$, which may admit an analytic solution.
In contrast to the trapped system, the homogeneous system has a subtle feature that impedes a direct approach to perturbation theory, namely, the non-vanishing density for $\xi\ll|\chi|$. 
We must instead consider a regularized Lagrangian density~\cite{Kivshar1995a} for the homogeneous NLSE   [Eq.~(\ref{Perturbationequation}) with $\gamma R(u)\equiv 0$], found by appealing to N\"others theorem:
 \EQ{\label{Lagrangian}
 \mathcal{L}&=&\frac{i}{2}\left(u^*\frac{\partial u}{\partial \tau}-u\frac{\partial u^*}{\partial \tau}\right)\left(1-\frac{1}{|u|^2}\right)-\frac{1}{2}\left|\frac{\partial u}{\partial \chi}\right|^2-\frac{1}{2}\left(|u|^2-1\right)^2.
}
 The $(1-1/|u|^2)$ factor regularizes the Lagrangian density to give a vanishing contribution as $\chi\to \pm \infty$, and thus provides a suitable, divergence-free, starting point for a perturbative treatment of the dissipative dark soliton.
By minimizing the action of the Lagrangian density (\ref{Lagrangian}) for the ansatz $u\equiv u(\chi,a_1,\dots,a_n)$, while allowing for a non-zero value of the perturbation, $\gamma R(u)$, one finds the Euler-Lagrange equations 
 \begin{equation}\label{Lagrangecond}
 \frac{\partial L}{\partial a_j}-\frac{d}{d\tau}\left(\frac{\partial L}{\partial \dot{a}_j}\right)=2\gamma \Re \left(\int_{-\infty}^\infty d\chi R^*(u)\frac{\partial u}{\partial a_j}\right),
 \end{equation} 
 where $L\equiv \int^\infty_{-\infty} \mathcal{L}\;d\chi$, and $\Re$ denotes the real part. For a given perturbation and choice of ansatz,~\eref{Lagrangecond} provides equations of motion for the collective variables describing the system.
 
The wavefunction of a dark soliton in a homogeneous system is 
\EQ{\label{Dsol}
\psi&=&\sqrt{n}\left(i\frac{v}{c}+\frac{\xi}{\xi_v}\tanh\left(\frac{x-vt}{\xi_v}\right)\right)e^{-i \mu t/\hbar}.}
where $\xi_v=\xi/\sqrt{1-v^2/c^2}$ is the soliton width.
Rescaling as above and dropping the rotating phase factor [consistent with (\ref{Perturbationequation})], the soliton wavefunction can be written as
$u(\chi,\tau)=i \nu+\kappa\tanh\left[\kappa(\chi - \nu\tau)\right],$
with $\nu= v/c$, and $\nu^2+\kappa^2=1$. The simplest ansatz for the perturbed soliton is then
\EQ{\label{singlesolitonansatz}
u(\chi,\tau)=\kappa(\tau)&\tanh\left[\delta(\tau)(\chi-x_s(\tau))\right]+i\nu(\tau),
}
with constraint $\nu^2+\kappa^2=1$. The functions $\kappa(\tau)$, $\delta(\tau)$ and $x_s(\tau)$ describe the soliton depth, inverse-width, and location.  
The Lagrangian arising from this ansatz is
$L=2dx_s/d\tau\left[\tan^{-1}(\kappa/\nu)-\nu\kappa\right]-(2/3)\left(\kappa^2\delta+\kappa^4/\delta\right)$.
Solving the Euler-Lagrange equations \eref{Lagrangecond} without the effects of dissipation and noise then gives 
$\delta=\kappa$, $d\kappa/d\tau=0$, and $dx_s/d\tau=\nu,$ as expected.
\begin{figure}[!t]
\begin{center}
\includegraphics[width=0.6\columnwidth]{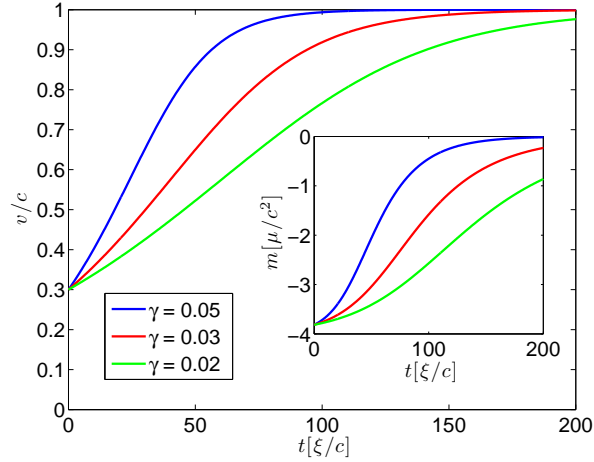}
\caption{(Color online) The velocity, (\ref{dissipativeresult}), and effective mass [(\ref{Solitonmassres}), inset] of a dark soliton in a homogenous quasi-one-dimensional Bose-Einstein condensate, for a range of damping rates. The initial velocity is $v_i=0.3c$.
\label{fig1}}
\end{center}
\end{figure}

\subsection{Soliton velocity}
We now solve for the evolution of the soliton velocity with dissipation. 
We consider the effects of the perturbation \eref{Perturbationterm} while neglecting the noise (setting $dw(\chi,\tau)\equiv 0$),
so that it reduces to
\begin{equation}\label{dissipation}
\gamma R(u)\equiv-i\gamma\left(-\frac{1}{2}\frac{\partial^2 u}{\partial \chi^2}+(|u|^2-1)u\right).
\end{equation}
We can expect this term to evolve a soliton with initial speed $\nu_i=v_i/c$ to $\nu_f=1$. 
Variational results for solitons under the effect of a number of types of dissipation have been obtained previously (see for example the review articles \cite{Kivshar1998a,Kivshar1989}, or \cite{Kivshar1995a, Theocharis2005, Kivshar1995a}).  However, an analytic treatment of dark soliton decay via the damped GPE defined by (\ref{Perturbationequation}) and (\ref{dissipation}) has not been previously provided.

The equations of motion for a single soliton in our dissipative system are given by \eref{Lagrangecond} as
\EQ{\label{1solitoneq1}
4\dot{x}_s\frac{\kappa^2}{\delta}&=&\frac{4}{3}\left(\kappa \delta+\frac{2\kappa^3}{\delta}\right),\\
\label{1solitoneq2}
4\dot{\nu}\kappa&=&\frac{8\gamma \nu \kappa^3}{3},
}
and $\kappa^2 = \delta^2$. Eq.~\eref{1solitoneq1} implies that $dx_s/d\tau=\pm \nu$, with $\kappa=\pm \delta$ respectively, corresponding to propagation in the positive or negative directions. Equation \eref{1solitoneq2}, with $\dot{\nu}=-\kappa\dot{\kappa}/\nu$, gives the soliton equation of motion
\EQ{\label{Seom}
\frac{d(\nu^2)}{d\tau}=\frac{4\gamma}{3} \nu^2(1-\nu^2),
}
with fixed points at $\nu=0$ and $\nu=\pm1$. Given initial speed $0<\nu_i<1$, the solution is
\begin{equation}\label{dissipativeresult}
\nu(\tau)=\frac{\nu_i}{\sqrt{(1-\nu_i^2)e^{-\frac{4\gamma }{3}\tau}+\nu_i^2}}.
\end{equation}
Clearly $\nu\to \nu_i$ as $\tau\to 0$, and $\nu\to 1$ as $\tau\to\infty$, with characteristic timescale $\sim\gamma^{-1}$, or, in dimensional units, $\sim\xi/\gamma c\equiv \tau_d$.
This result can be used to give an analytical description of the dark soliton evolution due to the effects of thermal damping, including the decay time and a number of other useful properties, providing a point of comparison for full stochastic simulations. As shown in Figure \ref{fig1}, a dark soliton subject to damping accelerates from $0<\nu_i<1$ to $\nu=1$, corresponding to a particle with negative effective mass. 
\begin{figure}[!t]
\begin{center}
\includegraphics[width=0.6\columnwidth]{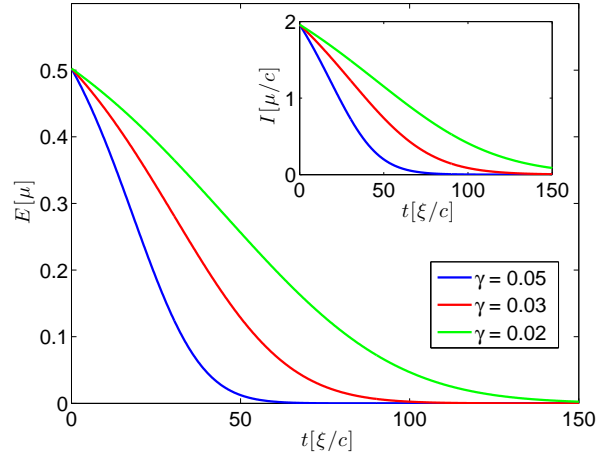}
\caption{(Color online) Energy and momentum (inset) as a function of time for a decaying dark soliton. The initial velocity is $v_i=0.3c$.
\label{fig2}}
\end{center}
\end{figure}

An immediate consequence of obtaining a damping solution is that we can easily evaluate the dark soliton energy $E = (4\mu/3)(1-\nu^2)^{3/2}$, mass $m = -(4\mu/c^2)\sqrt{1-\nu^2}$, and momentum $Ic/\mu=-2\nu\sqrt{1-\nu^2}+2\tan^{-1}\left(\sqrt{1-\nu^2}/\nu\right)$:
\EQ{\label{Solitonenergyres}
E(\tau)&=&E_i\left[1+\nu_i^2(e^{4\gamma \tau/3}-1)\right]^{-3/2},\\
\label{Solitonmassres}
m(\tau)&=&m_i \frac{e^{-\frac{2\gamma \tau}{3}}}{\sqrt{\left(1-\nu_i^2\right)e^{-\frac{4\gamma \tau}{3}}+\nu_i^2}},\\
\label{Solitonmomentumres}
\frac{I(\tau)c}{\mu}&=&\frac{-2\nu_i\sqrt{1-\nu_i^2}e^{-2\gamma \tau/3}}{(1-\nu_i^2)e^{-4\gamma \tau/3}+\nu_i^2}+2\tan^{-1}\left(\frac{\sqrt{1-\nu_i^2}}{\nu_i}e^{-2\gamma \tau/3}\right).
}
An example of the soliton mass is shown in Figure \ref{fig1} (inset), and the energy and momentum are shown in Figure \ref{fig2}. 
\begin{figure}[!t]
\begin{center}
\includegraphics[width=0.6\columnwidth]{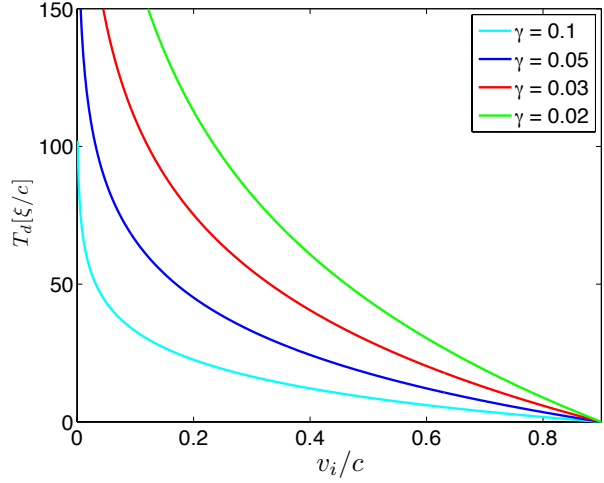}
\caption{(Color online) The decay time [Eq. \eref{lifetime}] of a damped soliton in a homogenous BEC, versus initial velocity $v_i$, for a soliton that is resolvable up to a velocity $v_f=0.9c$. \label{fig3}}
\end{center}
\end{figure}
\par
We can also define a decay-time $T_d$ for the soliton as the time for the velocity to evolve between $\nu_i$ and $\nu_f$. 
Using \eref{dissipativeresult} we find 
\EQ{\label{lifetime}
T_d&=&\frac{\xi}{c}\int_{\nu_i}^{\nu_f}d\nu\left(\frac{d\nu}{d\tau}\right)^{-1}=\frac{3\xi}{4\gamma c}\ln\left(\frac{\nu_f^2}{1-\nu_f^2}\frac{1-\nu_i^2}{\nu_i^2}\right).\;\;\;\;\;
}
The decay-time diverges for $\nu_i\to 0$ or $\nu_f\to 1$, but the two divergences have a different physical origin. When $\nu_i\to 0$ the soliton becomes immune to damping. The state $\nu=0$ is an eigenstate of parity. In practice any small thermal or quantum fluctuation will cause symmetry breaking and destabilize the $\nu_i=0$ state, forcing a direction of motion to be selected. The divergence as $\nu_f\to 1$ is due to the asymptotic nature of the decay. As the soliton evolves it will eventually become too shallow to be experimentally resolved, at which point the soliton has decayed for all practical purposes.  
The dependence of soliton decay time on $\nu_i$ is shown in Figure \ref{fig3} for a range of damping rates and $\nu_f=0.9$. The logarithmic dependence on velocities seen in \eref{lifetime} is qualitatively similar to that found by Fedichev {\em et al} (Eq. (13) of Ref.~\cite{Fedichev1999b}), via different reasoning. 

\begin{figure}
\begin{center}
\includegraphics[width=0.6\columnwidth]{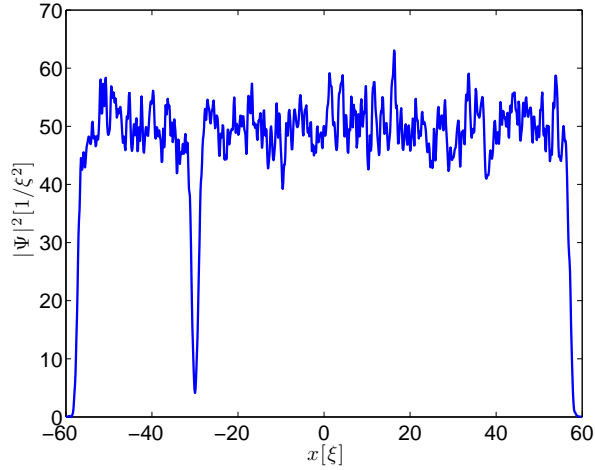}
\caption{A single soliton imprinted on a sample from the SGPE ensemble.  The external potential is zero within the condensate, and provides steep confinement near $\chi=\pm 60$. The temperature is $\sigma=0.3$.
\label{fig4}}
\end{center}
\end{figure}

\section{Stochastic simulations}\label{SIMS}
Our primary interest in this work is the influence of noise on the soliton velocity. 
In this section we numerically simulate the evolution of a dark soliton in a large, finite, one dimensional system, in order to assess the effect of thermal fluctuations on the soliton's velocity during its evolution. Our dimensionless stochastic equation of motion is 
\EQ{\label{spgpe}
du(\chi,\tau)&=&{\cal P}\Big\{(i-\gamma)(1-L_{GP})u(\chi,\tau)d\tau+dw(\chi,\tau)\Big\}
}
with $L_{GP}u=[-(1/2)\partial^2/\partial \chi^2+|u|^2+V(\chi)]u$, and complex-Gaussian noise defined by non-vanishing correlator
\EQ{\label{dimlessdw2}
\langle dw^*(\chi,\tau) dw(\chi',\tau')\rangle&=&2\gamma\sigma \delta(\chi'-\chi)\delta(\tau-\tau')d\tau.
}
The potential is chosen as $V(\chi)=0$ for $-60\lesssim\chi\lesssim 60$, with steep confinement at the endpoints, creating a homogeneous region of length $L_H\sim 120\xi$.
\begin{figure}[!t]
\begin{center}
\includegraphics[width=.6\columnwidth]{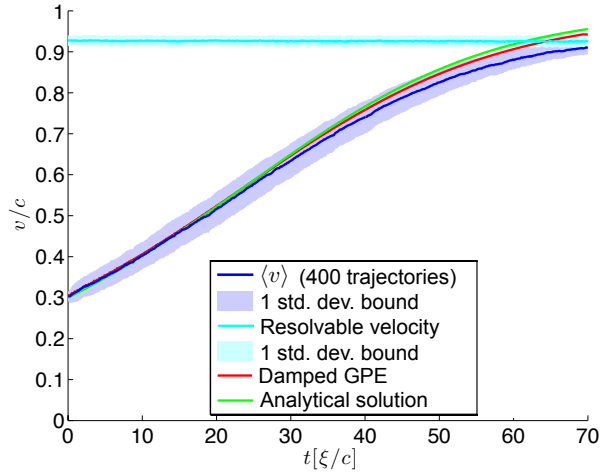}
\caption{(Color online) Ensemble-averaged velocity (extracted from soliton depth) of a dark soliton according to the SPGPE \eref{spgpe}, at temperature is $\sigma=0.1$. Also shown are the results from the damped GPE, and the analytical solution \eref{dissipativeresult}. The horizontal line shows the highest numerically resolvable velocity (which differs from 1 due to thermal fluctuations). 
\label{fig5}}
\end{center}
\end{figure}
We begin by creating a finite temperature equilibrium ensemble of trajectories for the confined system using the SPGPE. We initialize our dynamical simulations by imprinting a positively directed soliton wavefunction of given depth onto each trajectory. We use $v_i=0.3c$, and interaction parameter $g=0.02$ for all of our SPGPE simulations. For reasons of numerical expedience we also use $\gamma=0.05$. A detailed treatment of specific experimental systems requires that we include the full temperature dependence of $\gamma$. In this work we focus on the role of fluctuations, for fixed damping\footnote{However note that this choice of $\gamma$ is of the order of that realized in recent experiments~\cite{Blakie08a}.}. From Figure \ref{fig3} we can expect the soliton to damp out over a distance of order $\sim 40\xi$. We choose the initial position of the soliton as $x_s(0)=-30$ so that it remains well inside the homogeneous region throughout. Figure \ref{fig4} shows an element of the initial ensemble after the soliton has been imprinted.

We simulate the decay of a single soliton for 400 SGPE trajectories, and compare the results with damped PGPE simulations of a soliton with the same initial speed. To calculate the velocity of the soliton numerically, the minimum density of the condensate is found within the region $|\chi|<3L_H/8$, and the corresponding soliton velocity is calculated.  Eventually the soliton decays sufficiently that it becomes indistinguishable from the background fluctuations.  After this point the soliton is unresolvable, so we do not show soliton velocities after this time. 

\begin{figure}[!t]
\begin{center}
\includegraphics[width=.6\columnwidth]{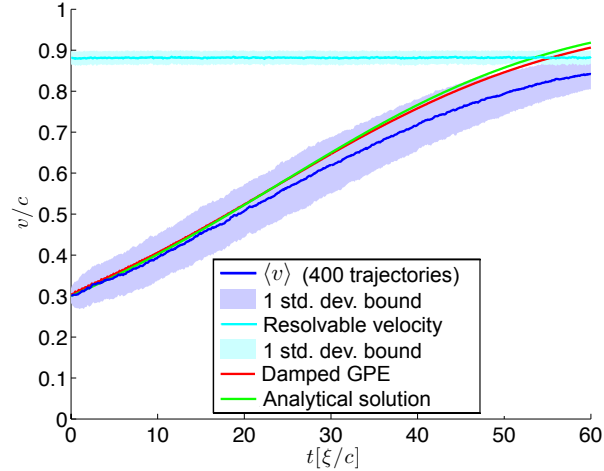}
\caption{(Color online) Velocity of a dark soliton according to the SPGPE \eref{spgpe} at temperature $\sigma=0.3$. All other parameters are the same as in Fig.~\ref{fig5}.
\label{fig6}}
\end{center}
\end{figure}
\begin{figure}[!t]
\begin{center}
\includegraphics[width=.6\columnwidth]{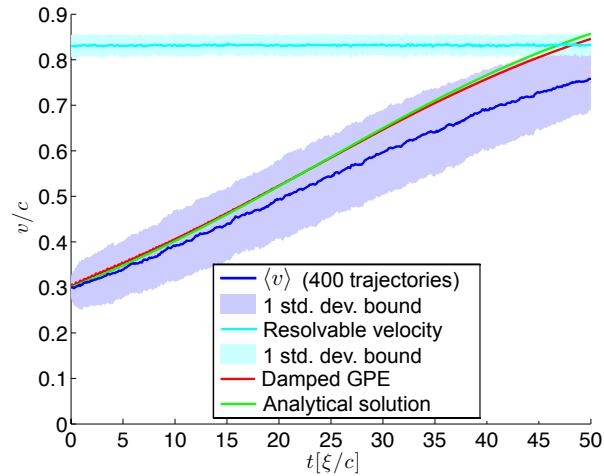}
\caption{(Color online) Velocity of a dark soliton according to the SPGPE \eref{spgpe} at temperature $\sigma=0.6$. All other parameters are the same as in Fig.~\ref{fig5}.
\label{fig7}}
\end{center}
\end{figure}

In figures \ref{fig5}, \ref{fig6}, and \ref{fig7} we compare the soliton velocity from the analytical result \eref{dissipativeresult}, with numerical simulations of the damped PGPE [\eref{Perturbationequation} and \eref{dissipation}], and with ensemble averages from \eref{spgpe}. The comparison is made for temperatures $\sigma=0.1$, $0.3$ and $0.6$. At low temperature [Figure \ref{fig5}], we see excellent agreement between the average SPGPE result and perturbation theory. At the higher temperatures the SPGPE shows a significant departure from both damped PGPE and perturbation theory. As temperature increases there is also a significant increase in the variance of the soliton velocity. Most notably, the stochastic fluctuations have the principal effect of slowing the decay process: the average soliton velocity in SPGPE is always lower than those found from damped GPE or soliton perturbation theory. 

Recent work on soliton core-filling \cite{Martin2010b} indicates that phase fluctuations in an ensemble of SPGPE trajectories should lower the expectation value of the soliton density and hence its instantaneous ensemble-averaged velocity. This expectation is confirmed by our analysis, and is a consequence of the classical relationship between the phase jump across the soliton ($\phi_c$) and its central density ($n_s$). In a system with background density $n_b$, the equation is $n_s=n_b\cos^2{(\phi_c/2)}$. Thermal fluctuations generate a temperature dependent variance of $\phi_c$ around the mean field solution. A higher temperature will thus yield a smaller value for $\langle n_s\rangle\propto \langle \cos^2{(\phi_c/2)}\rangle$.

Based on our observations of the stochastic slowing down of the a dark soliton, we infer that  \eref{lifetime} provides an analytical lower bound for the mean decay time of the dark soliton. 
\section{Conclusions}
In this work we have studied a novel regime for dark soliton dynamics in finite-temperature superfluids; this regime is characterized by temperatures comparable to the soliton energy, $k_B T\sim \mu$. We have established that in this regime the dark soliton exhibits an interesting physical phenomena: noise-induced slowing-down.

For the specific damping term arising in the high temperature classical field theory (the stochastic projected Gross-Pitaevskii equation, or SPGPE), we obtained an explicit solution for the velocity, and hence for the energy, mass and momentum, and decay time of a decaying dark soliton in a homogeneous BEC. Parity symmetry of the black soliton restricts these results to grey soliton initial conditions. We demonstrated the accuracy of the results at low temperature ($k_BT\ll \mu$) by comparing the soliton velocity with numerical simulations of the SPGPE, finding excellent agreement between the two approaches: in this regime the damping term dominates over the noise.
At high temperature ($k_BT\sim \mu$), the noise significantly modifies the soliton dynamics, and the analytical expression \eref{dissipativeresult} departs from numerical simulations of \eref{spgpe}. However, since the noise consistently \emph{slows} the decay of the soliton, our expression for the lifetime based on damping [Eq.~\eref{lifetime}] provides a lower bound for the dark soliton decay time. 

We conclude that a description of dissipation that includes damping \emph{and} the associated noise, leads to a dark soliton with increased stability. This contrasts strongly with the reduced stability of a quantized vortex in a spherically trapped 3D Bose gas~\cite{Rooney10a}, but is consistent with a recent low-temperature treatment  of the dark soliton~\cite{Martin2010b}.

Future work will focus on more detailed modeling of specific experimental systems and providing an analytical treatment of the noise. The latter requires treating noise in the collective variable equations of Lagrangian perturbation theory ---evidently a challenging task for the dark soliton. 

\ack We acknowledge A. G. Sykes, A. M. Martin, C. W. Gardiner for constructive interactions, and we would particularly like to acknowledge P. B. Blakie for many helpful discussions regarding this work. ASB acknowledges funding from New Zealand Foundation for Research, Science, and Technology contracts UOOX0801 and NERF-UOOX0703.
\section*{References}
\bibliographystyle{unsrt}


\begin{thebibliography}{10}

\bibitem{Burger99a}
S~Burger, K~Bongs, S~Dettmer, W~Ertmer, K~Sengstock, A~Sanpera, GV~Shlyapnikov,
  and M~Lewenstein.
\newblock {Dark solitons in Bose-Einstein condensates}.
\newblock {\em Phys. Rev. Lett.}, 83(25):5198--5201, 1999.

\bibitem{Strecker2002a}
K.~E. Strecker, G.~B. Partridge, A.~G. Truscott, and R.~G. Hulet.
\newblock Formation and propagation of matter-wave soliton trains.
\newblock {\em Nature}, 417(6885):150--153, 2002.

\bibitem{Kivshar1998a}
Y.~S. Kivshar and B.~Luther-Davies.
\newblock Dark optical solitons: physics and applications.
\newblock {\em Phys. Rep.}, 298(2-3):81--197, 1998.

\bibitem{Frantzeskakis10a}
D.~J Frantzeskakis.
\newblock {Dark solitons in atomic Bose-Einstein condensates: from theory to
  experiments}.
\newblock {\em J Phys A-Math Theor}, 43(21):213001, 2010.

\bibitem{Fetter2001}
A.~L. Fetter and A.~A. Svidzinsky.
\newblock {Vortices in a trapped dilute Bose-Einstein condensate}.
\newblock {\em J. Phys.: Condens. Matter}, 13:R135, 2001.

\bibitem{Denschlag2000a}
J.~Denschlag, J.~E. Simsarian, D.~L. Feder, C.~W. Clark, L.~A. Collins,
  J.~Cubizolles, L.~Deng, E.~W. Hagley, K.~Helmerson, W.~P. Reinhardt, S.~L.
  Rolston, B.~I. Schneider, and W.~D. Phillips.
\newblock {Generating solitons by phase engineering of a Bose-Einstein
  condensate}.
\newblock {\em Science}, 287(5450):97--101, 2000.

\bibitem{Anderson2001b}
B.~P. Anderson, P.~C. Haljan, C.~A. Regal, D.~L. Feder, L.~A. Collins, C.~W.
  Clark, and E.~A. Cornell.
\newblock {Watching dark solitons decay into vortex rings in a Bose-Einstein
  condensate}.
\newblock {\em Phys. Rev. Lett.}, 86(14):2926--2929, 2001.

\bibitem{Becker08a}
Christoph Becker, Simon Stellmer, Parvis Soltan-Panahi, Soeren Doerscher,
  Mathis Baumert, Eva-Maria Richter, Jochen Kronjaeger, Kai Bongs, and Klaus
  Sengstock.
\newblock {Oscillations and interactions of dark and dark-bright solitons in
  Bose-Einstein condensates}.
\newblock {\em Nat Phys}, 4(6):496--501, 2008.

\bibitem{Stellmer08a}
S~Stellmer, C~Becker, P~Soltan-Panahi, E.~M Richter, S~Doerscher, M~Baumert,
  J~Kronjaeger, K~Bongs, and K~Sengstock.
\newblock {Collisions of dark solitons in elongated Bose-Einstein condensates}.
\newblock {\em Phys. Rev. Lett.}, 101(12):120406, 2008.

\bibitem{Dutton2001a}
Z.~Dutton, M.~Budde, C.~Slowe, and L.~V. Hau.
\newblock {Observation of quantum schock waves created with ultra-compressed
  slow light pulses in a Bose-Einstein condensate}.
\newblock {\em {Science}}, 293:663, 2001.

\bibitem{Engels07a}
P.~Engels and C.~Atherton.
\newblock {Stationary and nonstationary fluid flow of a Bose-Einstein
  condensate through a penetrable barrier}.
\newblock {\em Phys. Rev. Lett.}, 99(16):160405, 2007.

\bibitem{Weller08a}
A~Weller, J.~P Ronzheimer, C~Gross, J~Esteve, M.~K Oberthaler, D.~J.
  Frantzeskakis, G~Theocharis, and P.~G Kevrekidis.
\newblock Experimental observation of oscillating and interacting matter wave
  dark solitons.
\newblock {\em Phys. Rev. Lett.}, 101(13):4, 2008.

\bibitem{Hamner11a}
C~Hamner, J.~J Chang, P~Engels, and M.~A Hoefer.
\newblock Generation of dark-bright soliton trains in superfluid-superfluid
  counterflow.
\newblock {\em Phys. Rev. Lett.}, 106(6):065302, 2011.

\bibitem{Fedichev1999b}
P.~O. Fedichev, A.~E. Muryshev, and G.~V. Shlyapnikov.
\newblock Dissipative dynamics of a kink state in a bose-condensed gas.
\newblock {\em Phys. Rev. A}, 60(4):3220--3224, 1999.

\bibitem{Busch2000}
T.~Busch and J.~R. Anglin.
\newblock {Motion of dark solitons in trapped Bose-Einstein condensates}.
\newblock {\em Phys. Rev. Lett.}, 84(11):2298--2301, 2000.

\bibitem{Jackson07a}
B~Jackson, N.~P Proukakis, and C.~F Barenghi.
\newblock {Dark-soliton dynamics in Bose-Einstein condensates at finite
  temperature}.
\newblock {\em Phys. Rev. A}, 75(5):051601, 2007.

\bibitem{Cockburn10a}
S.~P. Cockburn, H.~E. Nistazakis, T.~P. Horikis, P.~G. Kevrekidis, N.~P.
  Proukakis, and D.~J. Frantzeskakis.
\newblock Matter-wave dark solitons: Stochastic versus analytical results.
\newblock {\em Phys. Rev. Lett.}, 104(17):174101, Apr 2010.

\bibitem{Martin2010b}
A.~D Martin and J~Ruostekoski.
\newblock Nonequilibrium quantum dynamics of atomic dark solitons.
\newblock {\em New J. Phys.}, 12:055018, 2010.

\bibitem{Dziarmaga2003a}
J.~Dziarmaga, Z.~P. Karkuszewski, and K.~Sacha.
\newblock Images of the dark soliton in a depleted condensate.
\newblock {\em J Phys B-At Mol Opt}, 36(6):1217--1229, 2003.

\bibitem{Dziarmaga04a}
J.~Dziarmaga.
\newblock Quantum dark soliton: Nonperturbative diffusion of phase and
  position.
\newblock {\em Phys. Rev. A}, 70(6):063616, 2004.

\bibitem{Dziarmaga2006a}
J.~Dziarmaga and K.~Sacha.
\newblock {Images of a Bose-Einstein condensate: diagonal dynamical Bogoliubov
  vacuum}.
\newblock {\em J. Phys. B-At. Mol. Opt.}, 39(1):57--68, 2006.

\bibitem{Mishmash2009a}
R.~V. Mishmash and L.~D. Carr.
\newblock Quantum entangled dark solitons formed by ultracold atoms in optical
  lattices.
\newblock {\em Phys. Rev. Lett.}, 103(14):140403, 2009.

\bibitem{Mishmash2009b}
R.~V. Mishmash, I.~Danshita, C.~W. Clark, and L.~D. Carr.
\newblock Quantum many-body dynamics of dark solitons in optical lattices.
\newblock {\em Phys. Rev. A}, 80(5):053612, 2009.

\bibitem{Dziarmaga2010a}
J.~Dziarmaga, P.~Deuar, and K.~Sacha.
\newblock Comment on "quantum entangled dark solitons formed by ultracold atoms
  in optical lattices''.
\newblock {\em Phys. Rev. Lett.}, 105(1):018903, 2010.

\bibitem{Sykes2011a}
A.~G. Sykes.
\newblock {Exact solutions to the four Goldstone modes around a dark soliton of
  the nonlinear Schr\"{o}dinger equation}.
\newblock {\em J. Phys. A-Math. Theor.}, 44(13):135206, 2011.

\bibitem{Muryshev2002a}
A.~Muryshev, G.~V. Shlyapnikov, W.~Ertmer, K.~Sengstock, and M.~Lewenstein.
\newblock {Dynamics of dark solitons in elongated Bose-Einstein condensates}.
\newblock {\em Phys. Rev. Lett.}, 89(11):110401, 2002.

\bibitem{Bilas2005a}
N.~Bilas and N.~Pavloff.
\newblock {Propagation of a dark soliton in a disordered Bose-Einstein
  condensate}.
\newblock {\em Phys. Rev. Lett.}, 95(13):130403, 2005.

\bibitem{Carr2001a}
L.~D. Carr, J.~N. Kutz, and W.~P. Reinhardt.
\newblock {Stability of stationary states in the cubic nonlinear Schrodinger
  equation: Applications to the Bose-Einstein condensate}.
\newblock {\em Phys. Rev. E}, 63(6):066604, 2001.

\bibitem{Proukakis2004a}
N.~P. Proukakis, N.~G. Parker, C.~F. Barenghi, and C.~S. Adams.
\newblock {Parametric driving of dark solitons in atomic Bose-Einstein
  condensates}.
\newblock {\em Phys. Rev. Lett.}, 93(13):130408, 2004.

\bibitem{Parker2003}
N.~G. Parker, N.~P. Proukakis, M.~Leadbeater, and C.~S. Adams.
\newblock {Soliton-sound interactions in quasi-one-dimensional Bose-Einstein
  condensates}.
\newblock {\em Phys. Rev. Lett.}, 90(22):220401, 2003.

\bibitem{Konotop2004a}
V.~V. Konotop and L.~Pitaevskii.
\newblock Landau dynamics of a grey soliton in a trapped condensate.
\newblock {\em Phys. Rev. Lett.}, 93(24):240403, 2004.

\bibitem{Brazhnyi2006a}
V.~A. Brazhnyi, V.~V. Konotop, and L.~P. Pitaevskii.
\newblock Dark solitons as quasiparticles in trapped condensates.
\newblock {\em Phys. Rev. A}, 73(5):053601, 2006.

\bibitem{Martin2010a}
A.~D Martin and J~Ruostekoski.
\newblock Quantum and thermal effects of dark solitons in a one-dimensional
  bose gas.
\newblock {\em Phys. Rev. Lett.}, 104(19):194102, 2010.

\bibitem{Stoof1999}
H.~T.~C. Stoof.
\newblock {Coherent verses Incoherent Dynamics during Bose-Einstein
  Condensation in Atomic Gases}.
\newblock {\em J. Low Temp. Phys.}, 114:11, 1999.

\bibitem{Pitaevskii1958}
L.~P. Pitaevskii.
\newblock {\em Zh. Eksp. Teor. Fiz.}, 35:408, 1958.

\bibitem{Arnold06a}
A.~S. Arnold, C.~S. Garvie, and E.~Riis.
\newblock {Large magnetic storage ring for Bose-Einstein condensates}.
\newblock {\em Phys. Rev. A}, 73:041606, 2006.

\bibitem{Schnelle08a}
S.~K. Schnelle, E.~D. van Ooijen, M.~J. Davis, N.~R Heckenberg, and
  H.~Rubinsztein-Dunlop.
\newblock Versatile two-dimensional potentials for ultra-cold atoms.
\newblock {\em Optics Express}, 16:1405, 2008.

\bibitem{Henderson09a}
K~Henderson, C~Ryu, C~MacCormick, and M.~G Boshier.
\newblock Experimental demonstration of painting arbitrary and dynamic
  potentials for bose-einstein condensates.
\newblock {\em New J. Phys.}, 11:043030, 2009.

\bibitem{Sherlock11a}
B.~E. Sherlock, M.~Gildemeister, E.~Owen, E.~Nugent, and C.~J. Foot.
\newblock Time-averaged adiabatic ring potential for ultracold atoms.
\newblock {\em Phys. Rev. A}, 83(4):043408, 2011.

\bibitem{Kivshar1995a}
Y.~S. Kivshar and W.~Krolikowski.
\newblock Lagrangian approach for dark solitons.
\newblock {\em Opt. Commun.}, 114(3-4):353--362, 1995.

\bibitem{Kivshar1994a}
Yuri~S. Kivshar and X.~Yang.
\newblock Perturbation-induced dynamics of dark solitons.
\newblock {\em Phys. Rev. E}, 49(2):1657--1670, Feb 1994.

\bibitem{Kivshar1989}
Yuri~S. Kivshar and Boris~A. Malomed.
\newblock Dynamics of solitons in nearly integrable systems.
\newblock {\em Rev. Mod. Phys.}, 61(4):763--915, Oct 1989.

\bibitem{Theocharis2005}
G~Theocharis, P~Schmelcher, MK~Oberthaler, PG~Kevrekidis, and DJ~Frantzeskakis.
\newblock {Lagrangian approach to the dynamics of dark matter-wave solitons}.
\newblock {\em Phys. Rev. A}, {72}:023609, {2005}.

\bibitem{Kong2010}
Qian Kong, Q.~Wang, O.~Bang, and W.~Krolikowski.
\newblock {Analytical theory for the dark-soliton interaction in nonlocal
  nonlinear materials with an arbitrary degree of nonlocality}.
\newblock {\em {Phys. Rev. A}}, {82}:013826, {2010}.

\bibitem{SGPEI}
C.~W. Gardiner, J.~R. Anglin, and T.~I.~A. Fudge.
\newblock {The Stochastic Gross-Pitaevskii equation}.
\newblock {\em J. Phys. B: At. Mol. Opt. Phys.}, 35:1555, 2002.

\bibitem{SGPEII}
C.~W. Gardiner and M.~J. Davis.
\newblock {The Stochastic Gross-Pitaevskii equation:II}.
\newblock {\em J. Phys. B}, 36:4731, 2003.

\bibitem{Blakie08a}
P.~B. Blakie, A.~S. Bradley, M.~J. Davis, R.~J. Ballagh, and C.~W. Gardiner.
\newblock {Dynamics and statistical mechanics of ultra-cold Bose gases using
  c-field techniques}.
\newblock {\em Adv. in Phys.}, 57(5):363--455, 2008.

\bibitem{Bradley2005b}
A.~S. Bradley, P.~B. Blakie, and C.~W. Gardiner.
\newblock {Properties of the stochastic Gross-Pitaevskii equation: finite
  temperature Ehrenfest relations and the optimal plane wave representation}.
\newblock {\em J. Phys. B: At. Mol. Opt. Phys.}, 38:4259, 2005.

\bibitem{Bradley08a}
A.~S. Bradley, C.~W. Gardiner, and M.~J. Davis.
\newblock {Bose-Einstein condensation from a rotating thermal cloud: Vortex
  nucleation and lattice formation}.
\newblock {\em Phys. Rev. A}, 77(3):033616, 2008.

\bibitem{Stoof2001}
H.~T.~C. Stoof and M.~J. Bijlsma.
\newblock {Dynamics of Fluctuating Bose-Einstein Condensate}.
\newblock {\em J. Low Temp. Phys.}, 124:431, 2001.

\bibitem{Proukakis06a}
N.~P. Proukakis, J.~Schmiedmayer, and H.~T.~C. Stoof.
\newblock Quasicondensate growth on an atom chip.
\newblock {\em Phys. Rev. A}, 73:053603, 2006.

\bibitem{Proukakis08a}
N.~P. Proukakis and B.~Jackson.
\newblock Finite-temperature models of bose-einstein condensation.
\newblock {\em J. Phys. B-At. Mol. Opt.}, 41(20):203002, 2008.

\bibitem{Cockburn09a}
S.~P. Cockburn and N.~P. Proukakis.
\newblock {The stochastic Gross-Pitaevskii equation and some applications}.
\newblock {\em Laser Phys}, 19(4):558--570, 2009.

\bibitem{Weiler08a}
C.~N. Weiler, T.~W. Neely, D.~R. Scherer, A.~S. Bradley, M.~J. Davis, and B.~P.
  Anderson.
\newblock {Spontaneous vortices in the formation of Bose-Einstein condensates}.
\newblock {\em Nature}, 455(7215):948, 2008.

\bibitem{Rooney10a}
S.~J. Rooney, A.~S. Bradley, and P.~B. Blakie.
\newblock {Decay of a quantum vortex: Test of nonequilibrium theories for warm
  Bose-Einstein condensates}.
\newblock {\em Phys. Rev. A}, 81(2):023630, 2010.

\end{thebibliography}

\end{document}